\documentclass[prb,showpacs,twocolumn,amsmath,amssymb]{revtex4}

\usepackage{graphicx}

\newcommand{\pdagger}{{\phantom{\dagger}}}
\newcommand{\dt}{\Delta\tau}

\begin{document}

\title{Efficiency of quantum Monte Carlo impurity solvers for dynamical mean-field theory}

\author{N.~Bl\"umer}
\email{Nils.Bluemer@uni-mainz.de}
\affiliation{Institute of Physics, Johannes Gutenberg University, 55099 Mainz, Germany}

\date{\today}

  \begin{abstract}
    Since the inception of the dynamical mean-field theory,
    numerous numerical studies have relied on the Hirsch-Fye quantum
    Monte Carlo (HF-QMC) method for solving the associated impurity
    problem.  Recently developed continuous-time algorithms (CT-QMC)
    avoid the Trotter discretization error and allow for faster
    configuration updates, which makes them candidates for replacing
    HF-QMC. We demonstrate, however, that a state-of-the-art
    implementation of HF-QMC (with extrapolation of discretization
    $\Delta\tau\to 0$) is competitive with CT-QMC. A quantitative
    analysis of Trotter errors in HF-QMC estimates and of appropriate
    $\dt$ values is included.

  \end{abstract}
  \maketitle


\section{Introduction}
  A conventional starting point for the study of strongly correlated
  electron systems is the Hubbard model, which, in its single-band
  version, reads
  \begin{equation}\label{eq:Hubbard}
    H=-t\sum_{\langle ij\rangle, \sigma}\big(c_{i\sigma}^\dagger 
  c_{j\sigma}^\pdagger + \text{H.c.}\big) + U \sum_{i}n_{i\uparrow}n_{i\downarrow}\,.
  \end{equation}
  Unfortunately, numerical methods for its direct solution are either
  restricted to one-dimensional cases or suffer, in general, from severe
  finite-size errors and/or sign problems. Insight into the physics of
  higher-dimensional systems, thus, requires the use of additional
  approximations. The dynamical mean-field theory (DMFT) neglects
  inter-site correlations by assuming a momentum-independent
  self-energy; it becomes exact in the limit of infinite coordination
  number. The DMFT maps the lattice problem onto a single-impurity
  Anderson model (SIAM), supplemented by a self-consistency
  condition.\cite{Georges96} Its enormous success within the last 15
  years would not have been possible without the availability of
  controlled numerical solvers for (multi-orbital) SIAMs, in particular, 
  of the auxiliary-field Hirsch-Fye quantum Monte Carlo
  (HF-QMC) algorithm.\cite{Hirsch86} 

  The HF-QMC method discretizes the imaginary-time path integral into
  $\Lambda$ time slices of uniform width $\dt=\beta/\Lambda$ (at
  finite temperature $T=1/\beta$); a Hubbard-Stratonovich (HS)
  transformation replaces the electron-electron interaction at each
  time step by a binary auxiliary field which is sampled by standard
  Markov Monte Carlo (MC) techniques. The numerically costly part in
  each MC step is the update of the Green function ($\Lambda\times
  \Lambda$ matrix), which involves ${\cal O}(\Lambda^3)$ operations.
  HF-QMC is numerically exact only in the limit $\dt\to0$; raw
  results (for fixed $\dt$) contain the Trotter error, a statistical
  error (which decays as $N^{-1/2}$ for $N$ MC sweeps), and -- in the
  DMFT context -- a convergency error. Keeping $\dt$ fixed for
  constant accuracy, the computational cost increases as $T^{-3}$,
  which limits the temperature range accessible to HF-QMC. The usage
  of HF-QMC as a DMFT solver requires specialized techniques for
  Fourier transforming the time-discretized Green function; violations
  of causality observed in early implementations\cite{Georges96} may be
  avoided by stabilizing spline interpolations with analytic
  high-frequency
  expansions.\cite{BluemerKnecht,Oudovenko02,Bluemer05ab}

  While fundamentally different DMFT solvers such as exact
  diagonalization or renormalization group methods often yield very
  useful information (in particular for the one-band case), they are
  more complementary than general alternatives to HF-QMC, e.g., for
  obtaining ground state or low-frequency results. Thus, much effort
  is being devoted to improving and extending HF-QMC, e.g., by
  incorporating projections to the ground state\cite{Feldbacher04} and
  new HS decouplings for spin-flip terms in multi-band Hubbard models
  (which otherwise lead to sign problems).\cite{Sakai0406} Very
  recently, two novel quantum Monte Carlo approaches have been
  formulated for solving the impurity problem in continuous imaginary
  time (CT-QMC), which are based on a weak-coupling
  expansion\cite{Rubtsov05} and an expansion in the impurity
  hybridization,\cite{Werner06} respectively.  Each of these CT-QMC
  methods, which eliminate the Trotter error in HF-QMC, is a potential
  candidate for superseding HF-QMC as general-purpose
  finite-temperature DMFT solver.

  The urgent need for quantitative comparisons between the CT-QMC and
  HF-QMC finite-temperature algorithms was soon realized. However, a
  first analysis of relative efficiency\cite{Gull06} 
  suffers from using an inefficient HF-QMC-DMFT variant and
  neglects the essential step of extrapolating the HF-QMC results to
  the limit $\dt=0$. It is the purpose of this paper to show that a
  correction of these issues reverses the result: for the given test
  case of a half-filled single-band Hubbard model [Eq.\ (\ref{eq:Hubbard})]
  with semi-elliptic density of states at low temperatures and for
  fixed computing resources, extrapolated HF-QMC results reach or
  surpass the precision of the CT-QMC methods.

  \section{Comparisons at equal CPU time} 
  In the following, we will extend the performance comparison
  initiated by Gull {\it et al.}\cite{Gull06} for the case of the
  interaction being chosen equal to the full bandwidth, $U=W=4$; the
  scale corresponds to setting $t=1/\sqrt{Z}$ in Eq.\ (\ref{eq:Hubbard}) on
  the Bethe lattice in the limit of infinite coordination number $Z$.
  Unless noted, all CT-QMC results shown in the following have been
  extracted from Ref.\ \onlinecite{Gull06}. These will be compared
  with new HF-QMC results at equal CPU time: 140 CPU hours on
  common AMD Opteron compute nodes.  Specifically, in the case of
  CT-QMC, 20 iterations of 7 hours each have been performed serially
  on a single AMD Opteron 244 CPU.\cite{GullPriv} In the case of
  HF-QMC, the same fixed total CPU time has been split up among 
  runs for  5--9 different discretizations.  Keeping, for simplicity, the
  number of sweeps constant ($10^6$ per iteration), this was
  accomplished by adjusting the number of iterations in a range of
  4--8 (which appears a bit small for reliable error analysis, but
  sufficient for this comparison).
  The advantage for the HF-QMC simulations of being run on a mix of
  slightly faster processors\cite{fn:CPUs} should be almost offset by
  their overhead of parallel execution. 

\begin{figure}
    \includegraphics[width=\columnwidth]{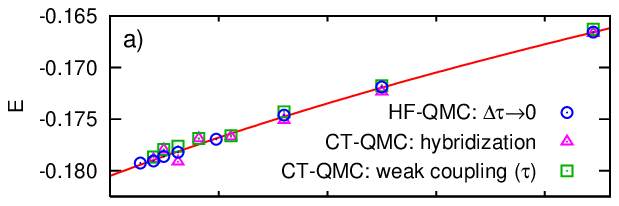}
    \includegraphics[width=\columnwidth]{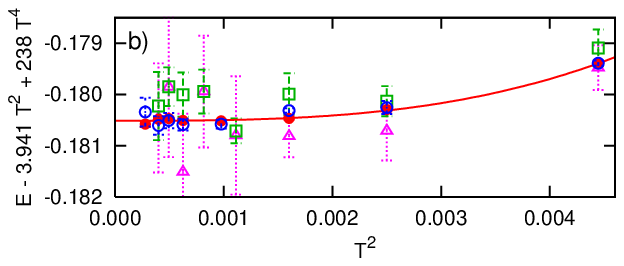}
    \caption{(Color online) a) Energy versus squared temperature: 
    extrapolated HF-QMC results (circles) and
      estimates (Ref.\ \protect\onlinecite{Gull06}) from CT-QMC
      (triangles, squares).  b) Data with leading $T$ corrections
      subtracted, plus reference (filled circles).}\label{fig:Etot}
  \end{figure}

  Figure \ref{fig:Etot}a 
  compares estimates of the energy, obtained from extrapolated HF-QMC,
  CT-QMC using the hybridization expansion,\cite{Werner06} and
  weak-coupling CT-QMC;\cite{Rubtsov05} also shown is a solid line
  which may be considered exact for the purpose of this comparison.
  Evidently, all QMC methods sustain reasonable accuracy down to
  fairly low temperatures $T\approx W/200$ with moderate computational
  cost. In order to resolve the differences, the
  leading temperature effects have been subtracted in Fig.\
  \ref{fig:Etot}b. At this scale, the CT-QMC results fluctuate visibly
  in a range consistent with their error bars, with possibly a slight
  positive bias for the weak-coupling variant.\cite{fn:tau} In
  contrast, the corresponding HF-QMC results show much smaller errors
  and are in excellent agreement with extreme-precision HF-QMC results
  (filled circles).

  For these reference results, 7 or 8 different discretizations using
  about 30--60 DMFT iterations with up to $5\cdot 10^6$ Monte Carlo
  sweeps each have been used for each temperature. All raw data points
  have carefully been tested for full convergence; due to the large
  number of iterations, autocorrelation times could be accurately
  determined. The HF-QMC runs for the lower temperatures have involved
  matrix sizes of up to $\Lambda_{\text{max}}=300$ with a computational
  cost of about 5000 CPU h per temperature; only for $T=1/40$
  with $\Lambda_{\text{max}}=320$ the effort was higher (about 15000
  CPU h). The fit curves to this data (solid lines in both panels
  of Fig.\ \ref{fig:Etot}) have the form
  \[
    E(T)=-0.18051 +   3.941\, T^2 -238\, T^4 + 12746\, T^6.
  \]
  Here, the leading finite-temperature correction $\tfrac{1}{2}\gamma
  T^2$ has been obtained from our estimate for the quasiparticle
  weight (at $T=0$, $U=4$) of $Z=0.2657\pm 0.0006$ via the relation
  $\gamma/2=\pi/(3 Z)$; only the other 3 coefficients were free fit
  parameters. For this reason, our ground state estimate
  $E_0=-0.18051\pm 0.00003$ has about the same (small) error bar as
  the best finite--temperature data.

\begin{figure}
    \includegraphics[width=\columnwidth]{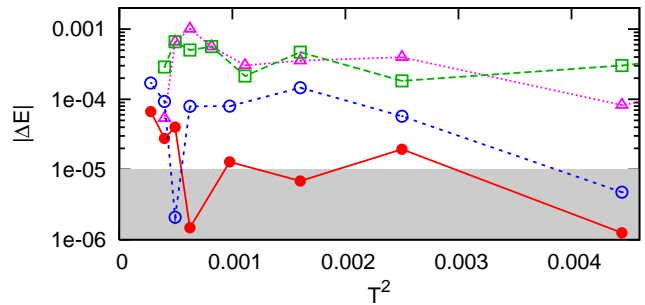}
    \caption{(Color online) Comparisons of errors in QMC energy
      estimates: absolute deviations of the data shown in Fig.\
      \ref{fig:Etot} from the reference curve on a logarithmic
      scale.}\label{fig:errors}
  \end{figure}

  A direct comparison of the accuracies of the different methods is
  shown in Fig.\ \ref{fig:errors}.
  The absolute errors (determined as deviation from the reference
  curve) of the hybridization CT-QMC results (triangles) generally
  increase from $10^{-4}$ at the highest temperatures to $10^{-3}$ at
  lower $T$; similar behavior (with stronger fluctuations) is observed
  for the weak-coupling CT-QMC data (squares). Already the ``regular''
  HF-QMC results (using 140 CPU h, open circles) are more
  accurate by nearly an order of magnitude, only to be surpassed by
  the more costly HF-QMC points (solid circles); note that all error
  estimates are reliable only down to the precision of the reference
  of about $10^{-5}$.  In the following, we will highlight the
  methodological ingredients that allow for the high
  demonstrated efficiency of our HF-QMC procedure.

  An essential feature of reliable HF-QMC studies is the inclusion of
  runs with different discretizations $\dt$ for estimating
  the impact of the associated systematic error; high-precision HF-QMC
  results can only be obtained by extrapolations $\dt\to0$ as
  illustrated in Fig.\ \ref{fig:Etau}a:
  \begin{figure}
    \includegraphics[width=\columnwidth]{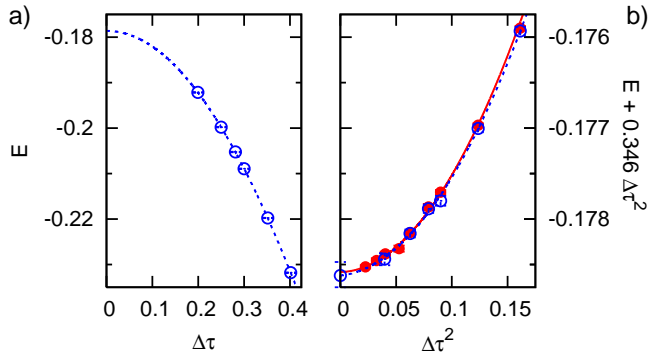}
    \caption{(Color online) a) HF-QMC estimate of energy at $T=1/45$
      (circles) versus discretization $\dt$ with extrapolation
      $\dt\to0$ (dashed line). b) Same data after subtraction of
      approximate leading $\dt$ error versus $\dt^2$ in comparison
      with reference HF-QMC data (filled circles and solid
      line).}\label{fig:Etau}
  \end{figure}
  The raw HF-QMC results for the energy (circles) vary by about $0.04$
  when the discretization is reduced from $\dt=0.4$ to $\dt=0.2$. As
  revealed by the extrapolation (dashed line), even the best raw
  results still contain systematic errors of order $10^{-2}$; however,
  the $\dt$ error is very regular and can essentially be eliminated in
  a quadratic least-squares fit in $\dt^2$ (cf.\ Sec.\ \ref{Trotter}).
  Already the subtraction of the (easily estimated) leading correction
  (which is linear in $\dt^2$) reduces the errors by one order of
  magnitude as shown in Fig.\ \ref{fig:Etau}b; only at this scale, the
  statistical error bars (of about $5\cdot 10^{-5}$) become visible.
  Full quadratic extrapolation (in $\dt^2$; dashed line) of the
  regular HF-QMC data results in a total error of $1.5\cdot
  10^{-4}$ (empty circle at $\dt^2=0$), i.e., in a reduction of the
  initial error by two full orders of magnitude. The excellent agreement
  with high-precision HF-QMC data (filled circles and solid line)
  confirms the validity of this procedure. Note that already the
  standard range of discretizations $\tau\in [0.2,0.4]$ in our HF-QMC
  data implies matrix sizes of up to 226 (at $T=1/45$), vastly greater
  than typical matrix sizes of about 80 for weak-coupling and about 12
  for hybridization expansion CT-QMC.\cite{Gull06}

\begin{figure}
    \includegraphics[width=\columnwidth]{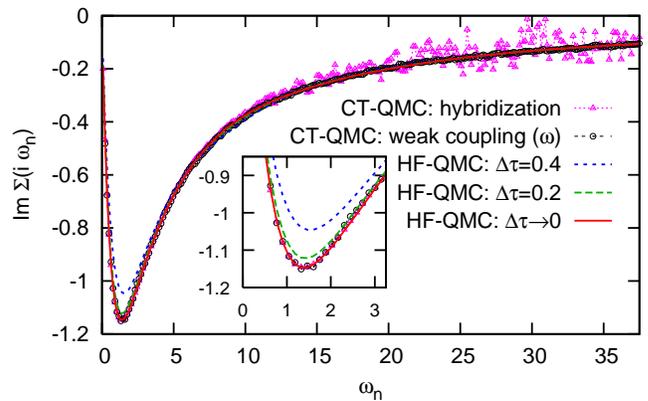}
    \caption{(Color online) Imaginary part of self-energy on the
      imaginary axis for $T=1/45$ as estimated by
      CT-QMC (Ref.\ \protect\onlinecite{Gull06}) using the hybridization expansion
      (triangles) or the weak-coupling expansion (circles) and by
      Hirsch-Fye QMC (dashed lines for $\dt=0.4, \dt=0.2$, solid line
      for $\dt\to 0$).}\label{fig:Sigma}
  \end{figure}

  An important prerequisite for extrapolation schemes such as
  described above is that all errors in the raw data (for fixed $\dt$)
  are well under control. By Fourier transforming only the difference
  between the QMC estimated Green function and a reference Green
  function (which is exact to order $U^2$), our HF-QMC
  implementation\cite{BluemerKnecht} eliminates $\dt$ errors beyond
  the inevitable Trotter contribution; at the same time,
  high-frequency fluctuations are reduced. This is  illustrated in
  the comparison of QMC self-energy estimates in Fig.\
  \ref{fig:Sigma}.
  Even the HF-QMC result for the coarsest discretization $\dt=0.4$
  (short-dashed line) deviates from the results for $\dt=0.2$ and
  $\dt\to 0$ (long-dashed and solid lines, respectively) visibly only
  for $\omega_n\lesssim 3$ (magnified in the inset). The extrapolated
  HF-QMC curve, in turn, agrees with the CT-QMC data at all
  frequencies, except for high-frequency fluctuations in the hybridization
  CT-QMC results. 
 
  As a consequence of the minimization of discretization errors in our
  HF-QMC implementation, also fluctuations in static quantities such
  as estimates of the kinetic energy are suppressed, as seen in Fig.\
  \ref{fig:Ekin_HF}:
  \begin{figure}
  \includegraphics[width=\columnwidth]{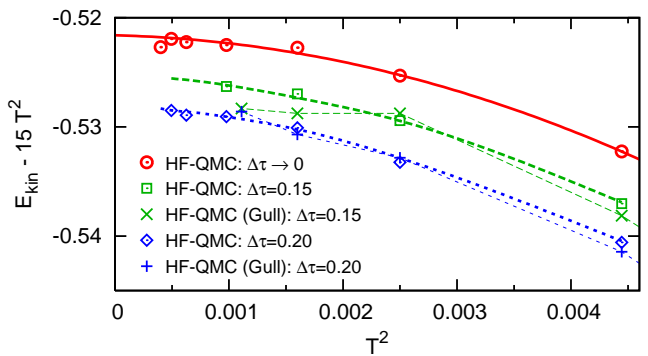}
  \caption{(Color online) Kinetic energy versus squared temperature (minus leading $T$ corrections):
    estimates from HF-QMC at finite discretization (squares for
    $\dt=0.15$, diamonds for $\dt=0.2$) and after extrapolation
    $\dt\to 0$ (circles); dashed and solid lines denote
    corresponding reference results. Also shown: HF-QMC results
    extracted from Fig.\ 3 in Ref.\ \onlinecite{Gull06} (crosses);
    thin lines are guides to the eye only.}\label{fig:Ekin_HF}
  \end{figure}
  our HF-QMC results at finite discretization (squares/diamonds)
  follow the reference results (dashed lines) with only small scatter;
  thus, a reliable extrapolation $\dt\to 0$ is possible (circles and
  solid line). In contrast, the HF-QMC results from Ref.\
  \onlinecite{Gull06} (crosses) show an order of magnitude larger
  fluctuations and do not allow for a precise extrapolation, although
  they have been obtained at higher cost (140 CPU h per data point)
  than each of our finite--$\dt$ results (for which between 5 and 70
  CPU h were used, depending on temperature). Evidently, the HF-QMC
  implementation used in Ref.\ \onlinecite{Gull06} is much less
  efficient.

\begin{figure}
  \includegraphics[width=\columnwidth]{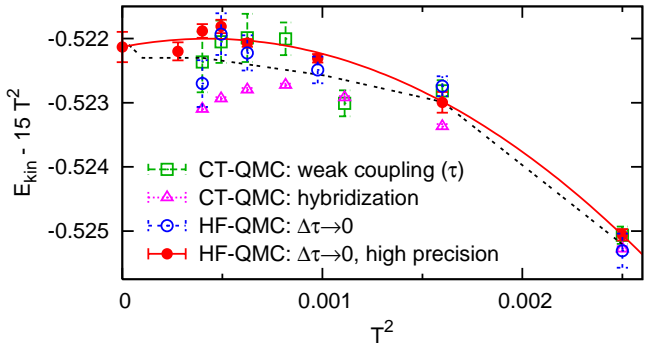}
  \caption{(Color online) Kinetic energy versus squared temperature: estimates at fixed CPU time from
    CT-QMC (Ref.\ \protect\onlinecite{Gull06})  (triangles, squares) and HF-QMC (empty
    circles) after subtracting leading $T$ corrections. Also shown are
    higher-precision results from HF-QMC (filled circles and solid
    line) and hybridization CT-QMC (double-dashed
    line).}\label{fig:Ekin}
  \end{figure}

  As seen in Fig.\ \ref{fig:Ekin},
  the estimates of the kinetic energy obtained from weak-coupling
  CT-QMC and $\dt$-extrapolated regular HF-QMC agree within their
  (similar) error bars with the reference HF-QMC results; the
  extrapolation of the latter for $T\to 0$ (solid line) is also in
  excellent agreement with low-$T$ hybridization CT-QMC data of Ref.\
  \onlinecite{Werner06} (double-dashed line). The regular
  hybridization CT-QMC data (triangles) shows by a factor of about 2
  larger deviations from the reference (this deviation is systematic
  and far beyond the stated error bars). So, all 3 QMC methods show
  roughly the same efficiency in estimating this observable.


  \section{Quantitative study of HF-QMC Trotter errors}\label{Trotter}
\begin{figure}
    \includegraphics[width=\columnwidth]{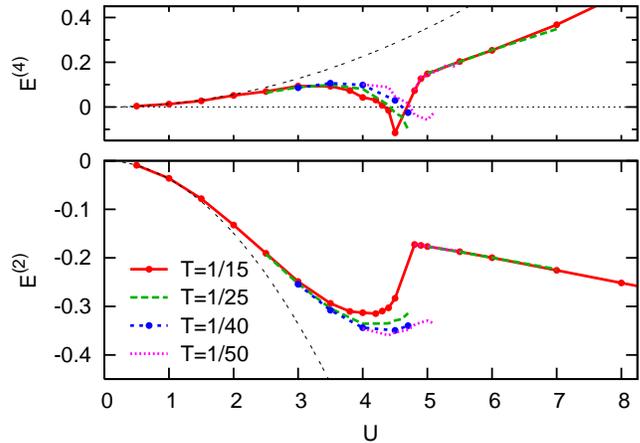}
    \caption{(Color online) Coefficients of Trotter errors in HF-QMC
      estimates of the energy (see text).}\label{fig:TrotterEtot}
  \end{figure}

  As we have established above for the specific test case defined in
  Ref.\ \onlinecite{Gull06}, HF-QMC results should be extrapolated to
  $\dt=0$, if possible. Many HF-QMC studies, however, include
  simulations only for a single value of $\dt$. The
  reliability of such studies depends on whether the
  chosen values of $\dt$ can be regarded as ``small''. 
  Up to now, many different criteria have been used in the literature,
  such as (for $t^*=1$) $\dt\lesssim 0.25$ or $\dt\lesssim 1/(5U)$
  etc.\cite{Werner06} 
  We have invested significant computing resources for
  determining the leading and subleading Trotter errors in HF-QMC
  estimates of selected observables throughout the (paramagnetic)
  phase space of model (\ref{eq:Hubbard}).
   Figure \ref{fig:TrotterEtot} 
  shows the coefficients in an expansion of the form
  \[
 E(U,\!T;\dt)\approx E^{(0)}(U,\!T) +  E^{(2)}(U,\!T) \dt^2 + E^{(4)}(U,\!T) \dt^4
  \]
  versus interaction $U$ for a range of temperatures extending from a
  value slightly above the critical Mott end point ($U\approx 4.67$,
  $T\approx 0.055$) to much smaller values. For weak interaction
  $U\lesssim 2$, both the leading correction $E^{(2)}$
  (lower panel) and the subleading term $E^{(4)}$ (upper
  panel) scale asymptotically as $U^2$ (double-dashed lines); thus,
  in this regime the Trotter error is indeed proportional
  to $(U\dt)^2$. For larger interactions, however, the error grows
  much slower; it even decreases as a function of $U$ in the strongly
  correlated metallic phase ($4\lesssim U\lesssim 5$). Only in this
  range, a temperature dependence of the coefficients becomes clearly
  visible. Finally, in the insulating phase ($U\gtrsim 5$), the 
  leading coefficient $E^{(2)}$ grows essentially
  linearly with $U$, so that $\dt\sqrt{U}$ would have to be kept
  constant for constant Trotter error.

\begin{figure}
    \includegraphics[width=\columnwidth]{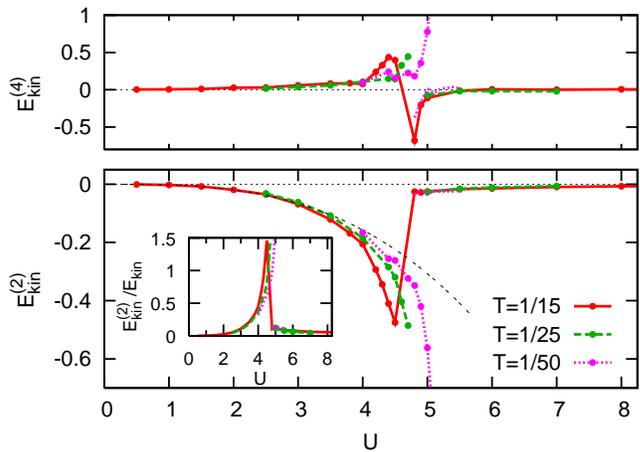}
    \caption{(Color online) Coefficients of Trotter errors in HF-QMC
      estimates of kinetic energy. Inset: relative coefficients.}\label{fig:TrotterEkin}
  \end{figure}

  The corresponding coefficients in the HF-QMC Trotter error expansion for the kinetic energy show markedly different behaviors in Fig.\ \ref{fig:TrotterEkin}.
  Here, the leading coefficient $E_{\text{kin}}^{(2)}$ scales as $U^3$
  for $U\lesssim 3$; $E_{\text{kin}}^{(4)}$ is too noisy for this kind
  of analysis. Both coefficients are strongly enhanced near the Mott
  transitions, with a pole-like structure in $E_{\text{kin}}^{(4)}$.
  Again, the temperature dependences are appreciable only in this
  range. Somewhat surprisingly, the Trotter errors decay quickly in
  the insulating phase, even the relative errors (see inset of Fig.\
  \ref{fig:TrotterEkin}); this observation may be specific to our
  implementation.

\begin{figure}
    \includegraphics[width=\columnwidth]{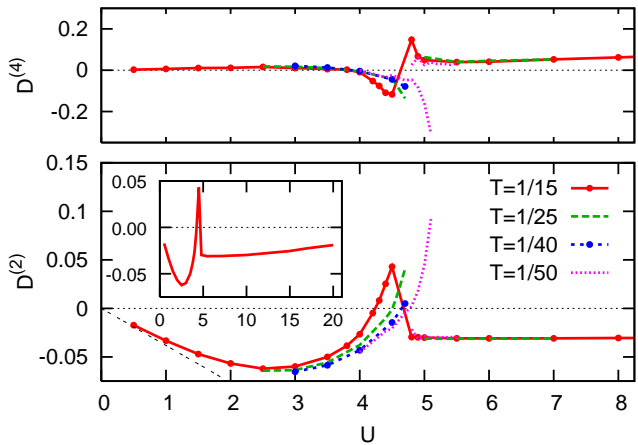}
    \caption{(Color online) Coefficients of Trotter errors in HF-QMC
      estimates of double occupancy.}\label{fig:TrotterD}
  \end{figure}

  In the case of the double occupancy $D=\langle
  n_{i\uparrow}n_{i\downarrow}\rangle$ (see Fig.\ \ref{fig:TrotterD}),
  the Trotter error starts off as $\dt^2 U$ for very small
  interactions $U\lesssim 1$. It then decays with a sign change just
  below the Mott transition. In the insulating phase, both
  coefficients are nearly constant; a decay of $D^{(2)}$ for extremely
  large $U$ is visible in the inset. 

  \section{ Conclusion} 
  We have compared published
  results\cite{Gull06} of two different continuous-time QMC methods
  with data obtained from a Hirsch-Fye QMC implementation by
  extrapolating the discretization $\dt\to0$.  HF-QMC involved typical
  matrix sizes more than twice as large as weak-coupling CT-QMC and
  nearly 20 times larger than hybridization CT-QMC (cf.\ Fig.\ 2 in
  Ref.\ \onlinecite{Gull06}). If this was the only relevant factor,
  HF-QMC should (due to the cubic scaling of the runtime of each
  method with the matrix size) be less efficient by one order of
  magnitude than the weak-coupling and by four orders of magnitude
  than the hybridization expansion variant.
  However, at (roughly) equal numerical effort, the HF-QMC results are
  about equally accurate for the kinetic energy and an order of
  magnitude more accurate for the energy;
  the latter translates\cite{fn:CLT} into two orders of
  magnitude higher efficiency of HF-QMC.  Apparently, much larger
  fluctuations in the CT-QMC methods outweigh the advantage of using
  smaller matrices (in the temperature range included in this
  comparison). An important factor is certainly the computation of
  Green functions which is (on the discretization grid) trivial for
  HF-QMC (with $\Lambda$ measurements for each time difference
  $\tau_i-\tau_j$ resulting from each sweep) while it requires
  significant additional efforts in both CT-QMC methods.\cite{Gull06}

  We have also quantified contributions to the Trotter error in HF-QMC
  estimates of energetics which show quite complex behavior. A rule
  for ``small enough'' values of $\dt$ which holds approximately for
  large $U$ can be obtained by requiring the quartic corrections (for
  $E$ or $D$) to be smaller than the quadratic ones: $\dt\ll
  1/\sqrt{0.28 U t^*}$. The observed small temperature dependence of
  the coefficients could be used for cheap high-precision studies at
  low $T$ (by extrapolating the coefficients and performing low-$T$
  simulations only for large $\dt$). A more surprising insight is that
  the Trotter errors in $D$ are unusually small for $U\approx 4.8$ and
  $T\lesssim 40$; the same holds true for $E^{(4)}$ so that HF-QMC is
  particularly precise in this range.  More generally, HF-QMC remains
  the method of choice for obtaining precise energetics, even
  extrapolated to the ground state.\cite{Bluemer05ab}

  The CT-QMC methods are certainly important and promising additions
  to the portfolio of DMFT impurity solvers. They allow for direct
  simulations at extremely low temperatures without requiring
  extrapolations in $\dt$ and/or $T$ and appear to offer greater
  flexibility in the choice of the Hamiltonian without incurring
  significant sign problems. However, as shown in this study, 
  HF-QMC (with extrapolation $\dt\to 0$) can be equally or even
  more efficient and should not be discarded (yet).

  \section*{Acknowledgments} 
  \vspace{-2ex} 
Useful discussions with F.\ Assaad, P.\ G.\ J.\ van
  Dongen, F.\ Gebhard, E.\ Gull, K.\ Held, A.\ Lichtenstein, D.
  Vollhardt, and P.\ Werner as well as code contributions by C. Knecht
  and support by the DFG (Forschergruppe 559, Bl775/1) are gratefully
  acknowledged.


\begin{thebibliography}{16}
\bibitem{fn:CPUs} The HF-QMC results have been obtained by parallel executions on 8 (sometimes 16) processor cores on a mix on AMD processors (Opteron 244, 246, 270, and 2216).
\bibitem{fn:tau} Here and in the following, we show the more favorable set of weak-coupling CT-QMC results: measurement in $\tau$ for $E_{\text{kin}}$, measurement in $\omega$ for $\Sigma(i\omega_n)$; cf. Ref.\ \onlinecite{Gull06}.
\end{thebibliography}
\end{document}